\documentclass[12pt]{article}
\title{Dynamics of QCD at large $N_c$\thanks{
Invited Talk at the International  Conference
on Problems of Quantum field Theory, Dubna, Russia,
July 1998}
}
\author{Yu.A.Simonov\\ Institute of Theoretical and Experimental
Physics\\ 117218, Moscow, B.Cheremushkinskaya 25,\\ Russia}

\date{}
\def\la{\mathrel{\mathpalette\fun <}}

\def\fun#1#2{\lower3.6pt\vbox{\baselineskip0pt\lineskip.9pt
\ialign{$\mathsurround=0pt#1\hfil
##\hfil$\crcr#2\crcr\sim\crcr}}}
\newcommand{\be}{\begin{equation}}
 \newcommand{\ee}{\end{equation}}

\begin{document}
\maketitle

\begin{abstract}

Dynamics of confinement, chiral symmetry breaking and thermal phase
transition  is considered at  large $N_c$. It is argued that these
phenomena are quantitatively well described within the Gaussian
stochastic model of the QCD vacuum. Selfconsistent equations are
written for the field correlators of the model, yielding important
connection between gluonic correlation length of the vacuum and the
string tension. Comparison to other approaches and experimental and
lattice data is given.

\end{abstract}

\section{Introduction}

Large $N_c$  limit first introduced for QCD in [1] has two
implications. First, it allows to neglect nonplanar perturbative
(P) diagrams and reduce nonpertubative (NP) background diagrams
to simple expressions; second, it allows to establish hierarchy of
physical characteristics, which can be compared to experiment.
E.g. the decay widths of all mesons $\Gamma_n$ are  $0(1/N_c)$ while
masses $M_n$ are $0(N_c^0)$, in experiment on average
$\Gamma_n/M_n\sim 0.1$ which gives an idea of the parameter for the
real QCD. Also white objects do not interact in the leading $N_c$
order, which leads to the so-called topological expansion of
high--energy scattering amplitudes. All dynamical picture of QCD
at large $N_c$ looks selfconsistent and realistic, and there have
been many attempts to derive it from the first principles [2].

Recently a new and universal method was suggested in QCD [3] which
allows to calculate all  amplitudes in terms of  a set of basic
quantities: field correlators (FC). The simplest approximation uses
only the lowest FC, quadratic in field strength, and was called the
Gaussian Stochastic Model (GSM). Both lattice calculations [4] and
theoretical estimates [5] show that accuracy of GSM is of the same
order as of the large $N_c$ expansion, and that GSM yields the
dominant contribution to the most QCD phenomena in the leading large
$N_c$ order.

The dynamical input of GSM -- the quadratic FC -- is usually taken in
a simple exponential form from the lattice data [4]. In the most
recent paper [6] the large $N_c$ limit was used to obtain selfcoupled
equations for FC, which in principle allow for selfconsistent
determination of FC from the QCD Lagrangian.  Keeping only the lowest
FC in these equations one is able to connect NP parameters ( string
tension, gluon condensate, correlation length) between themselves and
to the P parameter ($\Lambda_{QCD}$). In this way there appears a
realistic possibility to have a unified consistent picture of QCD
dynamics at large $N_c$, described by the  only scale parameter as it
should be. In this talk we discuss confinement (section 2), chiral
symmetry breaking (CSB) (section 3), thermal phase  transition
(section 4),  with concluding
remarks in section 5.

\section{Confinement at large $N_c$}

Lattice calculations performed at $R\la 1.5 fm$ confirm that static
fundamental quarks are confined by the linear potential $V(R)=\sigma
R$ [7] even in the presence of dynamical fermions [8] which in
principle should screen the quark charges at large distances. Since
dynamical quarks break the fundamental string in the next $0(1/N_c)$
order this means that in the region $R\la 1.5 fm$ the fundamental
string does not break and the leading $0(N^0_c)$ approximation is
valid.

Another feature of the confinement, also seen in lattice calculations
at $R\la 1.5 fm$ [7] is the so-called Casimir scaling [9], namely that
the string tension $\sigma (j)$ for static charges in the $j$
representation of the color group SU($N_c$) obey the law
\be
\frac{\sigma(j)}{\sigma{(fund)}}=\frac{C_2(j)}{C_2(fund)},
\ee
where  $C_2(j)$ is the quadratic Casimir operator,
\be
C_2(adj,N_c)=N_c,~~
C_2(fund,N_c)=\frac{N_c^2-1}{2N_c}.
\ee

The breaking of the adjoint string is the $0(1/N_c^2)$ process which
happens at large distance $R_{br}(N_c)$ which grows with $N_c$, as
can be seen from the expression [9]
\be
<W_{adj}(C)>=C_1exp(-\sigma_{adj} R T)+\frac{C_2}{N_c^2} exp
(-V_s(R)T).
\ee
Here $V_s(R) $ is the screened quark potential.

We now show that all the above features are nicely described in the
GSM. To this end one writes the nonabelian Stokes theorem [5,10]
 \be <W(C)>=<\frac{1}{N_C} P tr ~exp~ig~ \int_c A_{\mu}
dx_{\mu}>= \ee $$\frac{1}{N_c}<P ~tr~exp~ig~\int_S
d\sigma_{\mu\nu}F_{\mu\nu}(u,z_0)> $$ where we have defined \be
F_{\mu\nu}(u,z_0) = \Phi(z_0,u) F_{\mu\nu} (u) \Phi(u,z_0), \Phi(x,y)
= P \exp ig \int^x_y A_{\mu} dz_{\mu}
\ee
 and integration in (4) is
over the surface $S$ inside the contour $C$, while $z_0$ is an
arbitrary point, on which $<W(C)>$ evidently does not depend.  In the
Abelian case the parallel transporters $\Phi(z_0,u)$  and
 $\Phi(u,z_0)$ cancel and one obtains the usual Stokes theorem.

Note that the nonabelian Stokes theorem, eq. (4), is gauge invariant
even before averaging over all vacuum configurations -- the latter is
implied by the angular brackets in (4).

One can now use the cluster expansion theorem  to express the
r.h.s. of (4) in terms of $FC$, namely [3,5]
\be
<W(C)> = \frac{tr}{N_C} \exp \sum^{\infty}_{n=1} \frac{(ig)^n}{n!}
\int d\sigma(1)d\sigma(2)... d\sigma(n)\ll F(1)... F(n)\gg
\ee
where lower indices of $d\sigma_{\mu\nu}$ and $F_{\mu\nu}$ are
suppressed and $F(k) \equiv F_{\mu_k \nu_k}(u^{(k)}, z_0)$.

Note an important simplification -- the averages $\ll F(1) ...
F(n)\gg$ in the color symmetric vacuum are proportional to the unit
matrix in color space, and the ordering operator $P$ is not needed
any more.

Eq. (6) expresses Wilson loop in terms of gauge invariant $FC$,
also called cumulants , defined in terms of $FC$ as follows:\be
\ll F(1) F(2) \gg = < F(1)F(2)> - <F(1)><F(2)>
\ee
$$
\ll F(1) F(2) F(3) \gg = <F(1)F(2)F(3)>-\ll F(1)F(2)\gg
<F(3)>-
$$
$$
-<F(1)>\ll F(2)F(3)\gg - <F(2)>\ll F(1)F(3)\gg-<F(1)><F(2)><F(3)>
$$

In the lowest approximation, which corresponds to the GSM, one keeps
only the quadratic in F term, namely

\be
D_{\mu\nu\lambda\sigma} \equiv \frac {1}{N_c}
tr <F_{\mu\nu}(x) \Phi(x,y)F_{\lambda\sigma}(y)\Phi(y,x)>
\ee
The form (8) has a general decomposition in terms
of two Lorentz scalar functions $D(x-y)$ and $D_1(x-y)$ [3]
 \begin{eqnarray}
D_{\mu\nu\lambda\sigma} &=&
(\delta_{\mu\lambda}\delta_{\nu\lambda} -
\delta_{\mu\sigma}\delta_{\nu\lambda}){\cal{D}}(x-y) + \\
\nonumber
&+& \frac{1}{2} \partial_{\mu} \{[(h_{\lambda} \cdot
\delta_{\nu\sigma} - h_{\sigma}\delta_{\nu\lambda}) + ... ]
{\cal{D}}_1(x-y)\}
\end{eqnarray}
 Here the ellipsis implies terms obtained by permutation of indices.
 It is important that the second
term on the r.h.s. of (9) is a full derivative by construction.

Insertion of (9) into (6)  yields the area law of
Wilson loop with the string tension $\sigma$ $$<W(C)>= exp (-\sigma
S_{min})$$
 \be
  \sigma= \frac{1}{2}\int D(x)d^2x(1+O(FT_g^2))
  \ee
  where $O(FT_g^2)$
stands for the contribution of higher cumulants, and $S_{min}$ is the
minimal area for contour $C$.

The string tension for heavy (static) quarks is an  infinite sum of
FC of the field $F_{14}\equiv E_1$ integrated over the plane $14$:
$$
\sigma \sim \sum_n\frac{g^n}{n!}\prod^{n-1}_i d^2r_i\ll E_1(0)
E_1(r_1)E_1(r_1+r_2)...E_1(\sum r)\gg
$$
One can identify parameter of expansion in the sum above to be (only
even  powers of $n$ enter the sum)
$$
\zeta = (\bar E_1 T^2_g)^2
$$
where $T_g$ is the gluonic correlation length in the vacuum  and
$\bar E^2_1 \cong g^2<(E^a_1)^2>= \frac{4\pi^2}{12} G_2 \approx (0.2
GeV)^2
$ where $G_2$ is the standard gluon condensate.  Naively one would
expect that the correlation length $T_g$ is of the order of
confinement radius $R_c$, $T_g\sim R_c\sim \Lambda^{-1}_{QCD}$, in
which case since $\bar E_1\sim \Lambda^2_{QCD}$ the parameter $\zeta$
is $\zeta\sim 1$ and all FC are equally important (that
would be true for ordinary FC, but connected FC may have additional
small parameter at large $n$ due to cancellation in (7)).

However lattice calculations confirm that $T_g$ is much smaller [4],
indeed $T_g\approx 0.2 \div 0.3 fm$ and therefore parameter $\zeta$
is small
$$
\zeta=0.04 \div 0.1
$$

The regime  $\zeta\ll1, $ which seems to be characteristic for real
QCD, can be called the regime of the \underline{weak~~confinement}.
In this case the dynamics of quarks and gluons is adequately
described in most cases by the lowest (Gaussian) correlator [3,4].

Note that $D_1$ does not enter $\sigma$, but gives rise to the
perimeter  term and higher order curvature terms. On the other
hand  the lowest order perturbative QCD contributes to $D_1$ and not
to $D$, namely the one--gluon--exchange  contribution is
\be
D_1^{pert}(x)=\frac{16 \alpha_s}{3\pi x^4}
\ee

 Nonperturbative parts of $D(x)$ and $D_1(x)$ have been
 computed on the lattice [4]
 using the cooling method, which suppresses perturbative
 fluctuations. As one can see both functions are well
 described by an  exponent in the measured region, and $D_1(x)\sim \frac{1}{3}
 D(x)\sim exp (-x/T_g)$ , where
 $T_g\sim 0.2 fm$.

The string tension (10) can be  computed from the lattice data [4]
extrapolated to small distances, and agrees within 10-20\% with the standard
value $\sigma\approx 0.2 GeV^2$. Hence the Gaussian correlator alone can
explain the string tension.

We conclude this chapter with discussion of confinement for charges in
higher representations. As it was stated in the previous chapter, our
definition of confinement based on lattice data, predicts the
linear potential between static charges in any representation, with
string tension proportional to the quadratic Casimir operator.

Consider therefore the Wilson loop (4) for the charge in some
representation; the latter was not specified above in all eqs.
leading  to (10). One can write in general
\be
A_{\mu}(x)=A_{\mu}^aT^a, tr(T^aT^b)=\frac{1}{2}
\delta^{ab}
\ee
Similarly to (6) one has for the representation $j=(m_1,m_2...)$ of
the group $SU(N)$ with dimension $N(j)$
 \be <W(C)=\frac{1}{N(j)}tr_j
exp \sum^{\infty}_{n=1} \frac{(ig)^n}{n!} \int
d\sigma(1)...d\sigma(n)\ll F(1)...F(n)\gg \ee and by the usual
  arguments one has Eq.(9) .

              Due to the color neutrality of the vacuum each cumulant
              is proportional to the unit matrix in the color space,
e.g. for the lowest cumulant one has
\be
<F(1)F(2)>_{ab}=<F^c(1)F^d(2)>T^c_{an}T^d_{nb}=
\ee
$$
=<F^e(1)F^e(2)>\frac{1}{N^2_c-1}
T^c_{an}T^c_{nb}=\Lambda^{(2)} C_2(j)\cdot
\hat{1}_{ab},
$$
where we have used the definition
\be
T^cT^c=C_2(j)\hat{1}
\ee
and introduced a constant not depending on representation,
\be
\Lambda^{(2)}\equiv \frac{1}{N^2_C-1}<F^e(1)F^e(2)>,
\ee
and also used the color neutrality of the vacuum,
\be
<F^c(1)F^d(2)>=\delta_{cd}\frac{<F^e(1)F^e(2)>}{N^2_C-1}
\ee
For the next -- quartic cumulant one has
\be
\ll F(1)F(2)F(3)F(4)\gg_{\alpha\varepsilon}=
\ll F^{a_1}(1)F^{a_2}(2)F^{a_3}(3)F^{a_4}(4)\gg \times
\ee
$$
\times
T^{a_1}_{\alpha\beta}T^{a_2}_{\beta\gamma}T^{a_3}_{\gamma\delta}
T^{a_4}_{\delta\varepsilon}=
\Lambda^{(4)}_1(C_2(j))^2\delta_{\alpha\varepsilon}+\Lambda_2^{(4)}
(T^{a_1}T^{a_2} T^{a_1}T^{a_2})_{\alpha\varepsilon}
$$
Thus one  can see in the quartic cumulant a higher order of quadratic
Casimir and higher Casimir operators.

The string tension for the representation $j$ is the coefficient of
the diagonal element in (14) and (18)
\be
\sigma(j)=C_2(j)\int\frac{g^2\Lambda^{(2)}}{42}d^2x+O(C^2_2(j))
\ee
where the term $O(C^2_2(j))$ contains higher degrees of $C_2(j)$ and
higher Casimir operators.

Comparing our result (19) with lattice data [7]  one can
see that the first quadratic cumulant should be dominant as it
ensures proportionality of $\sigma(j)$ to the quadratic Casimir
operator.

Thus we see that GSM yields a simple confinement picture which is consistent
with lattice data and large $N_c$ considerations at least in the region $R\la
1.5 fm$. At larger $R$ and  fixed $N_c$ and for adjoint charges the screening
occurs which needs higher cumulants, and this happens in the $0(1/N_c^2)$
order.

\section{CSB at large $N_c$}

In the large $N_c$ limit the phenomenon of CSB is supported by the
Coleman--Witten theorem [11], whereas at $N_c=2,3$ lattice data show clearly
CSB in the order parameters of quark condensate $<\bar \psi (0) \psi (0)>$,
which is nonzero for $T\leq T_c$. Also parity doublets are missing in
the hadronic spectrum. There is still another important feature of CSB
seen in the heavy--light $q\bar Q$ system, namely the scalar confining
interaction for the light quark, which clearly  signals CSB. In the
present section we shall present results of recent studies of this system
[12,13], which is the simplest from the dynamical point of view.

We start with the quark Green's function and write the effective quark
Lagrangian in presence of a static source, using the  averaging of  the
partition function over gluonic field $A_\mu$.

To take into account the static source we consider the generalized coordinate
gauge [10] and express $A_\mu$ through $F_{\mu\nu}$ as
\be
A_\mu (x)=\int_C ds\frac{dz_{\alpha}(s,x)}{ds} F_{\alpha\beta} (z)
\frac{dz_{\beta}}{dx_\mu}
 \ee
where the contour $C$ starts at  $x_\mu$ and is described by $z_\mu(x,s)$
(in the usual coordinate gauge $z_\mu(x,s)=sx_\mu, 0\leq s\leq 1$).
The effective Lagrangian is (a more extended version of this derivation see
in [12]).
$$
{\cal{L}}_{eff}(\psi^+\psi)=\int \psi^+(x)(-i\hat \partial -im )
\psi(x)d^4x +
$$
\be
\frac{1}{2N_c}\int
d^4xd^4y(\psi^+_a(x)\gamma_{\mu}\psi_b(x))(\psi^+_b(y)
\gamma_{\mu'}\psi_a(y))\times
\ee
$$
\times
J_{\mu\mu'}(x,y)
$$
where we have defined
$$
J_{\mu\mu'}(z,w)=\int^z_C du_\alpha\int^w_Cdv_{\gamma}
(\delta_{\alpha\gamma}
\delta_{\beta\delta}-\delta_{\alpha\delta}
\delta_{\beta\gamma})
\frac{du_{\beta}}{dz_{\mu}}\frac{dv_{\delta}}{dw_{\mu'}}\times
$$
\be
\times D(u-v)
\ee
and $D(u)$ is
 defined in (9)
yielding the string tension
   \be
   \sigma = \frac{1}{2} \int^{\infty}_{ -\infty} d^2uD(u)
   \ee

Note that we have neglected in (21) higher field correlators, which
were argued above to yield subdominant contribution.

The Lagrangian (21) can be used to obtain equations for the quark Green's
function $S$ in the large $N_c$ limit, where the following rule of
replacement holds
\be
\psi_b(x)\psi^+_b(y)\to<\psi_b(x)\psi^+_b(y)>=N_cS(x,y),
\ee
One obtains a system of equations for the quark Green's function  $S$ and the
mass operator $M$
\be
iM(z,w)=J_{\mu\nu}(z,w)\gamma_{\mu}S(z,w)\gamma_{\nu}
\ee
\be
(-i\hat{\partial}_z-im)S(z,w)-i\int M(z,z')S(z', w)d^4z'=
\delta^{(4)}(z-w)
\ee

The system of equations (25,26)
is exact in the large $N_c$ limit, when higher correlators are neglected and
defines unambiguously both the interaction kernel $M$ and the Green's
function $S$.  One should stress at this point again that both $S$ and $M$
are not the one-particle operators but rather two--particle operators, with
the role of the second particle played by the static source. It is due to
this property, that $S$ and $M$ are gauge invariant operators, which is very
important to take confinement into account properly. Had we worked with
one--particle operators, as is the habit in QED and sometimes also in QCD,
then we would immediately loose the gauge invariance and the string, and
hence confinement.

The CSB can manifest itself in solutions of (25,26) in several ways. One is
the appearance of the nonzero chiral condensate
\be
<\bar \psi(0) \psi(0)>= iN_c tr~S(0,0)
\ee
This was estimated using the relativistic WKB method in [12] to be
\be
<\bar \psi (o) \psi(0)>\sim - N_c f(\frac{\sigma}{T_g}, \sigma^{3/2})
 \ee
  where $T_g$ is the gluonic correlation length [3,4].

Another manifestation of CSB is the scalar confinement, which is seen at
large distances $r$. Indeed one write expansion for the Green's function $S$
and $M$ in the inverse powers of the string mass $M_{str}=\sigma r+m$, and
the time--averaged Green's function $\bar S$ satisfies an equation [13]:
\be
[-i\vec \gamma\vec\partial -i(m+\sigma|\vec z|)]\bar S(\vec z, \vec w)=
\delta^{(3)}(\vec z-\vec w)
\ee

In (29) the string term $\sigma r$ enters as a scalar, which signals CSB.

Equations (28) and (29) exemplify the connection between CSB and confinement
in the large $N_c$ limit. The necessary appearance of CSB in this limit was
proved earlier in [11] but no hint to the possible mechanism was given. Here
we demonstrate that CSB occurs due to the string formation (which is
contained in factor $\sigma$ in (28), (29) and the kernel $J$ in (25),(26)),
but this effect comes to the existence only due to the solution of nonlinear
equations (25),(26). One should mention that these equations in
contrast to those of NJL,  are nonlocal and therefore do not need
cut-off.  The CSB solution is not obtained by perturbation expansion
of (25),(26), but rather is an extra, nonperturbative solution
existing due to the nonlinearity. A similar situation occurs in the
NJL model.

\section{Phase transition at large $N_c$}

To describe the phase  transition in QCD we shall use the basic idea,
suggested in [14], that the confining phase is governed by the NP fields,
while in the deconfining phase one has P fields in the NP background of
magnetic fields. Therefore one should first introduce the  background
formalism where both P and  NP  fields enter.

 We derive here basic formulas for the partition function, free energy and
Green's  function in the NP background formalism at
$T>0$ [15]. The total gluonic field $A_{\mu}$ is split into a
perturbative part $a_{\mu}$ and NP background $B_{\mu}$
\be
A_{\mu}=B_{\mu}+a_{\mu}
\ee
where both $B_{\mu}$ and  $a_{\mu}$ are subject to periodic boundary
conditions. The principle of this separation is immaterial for our purposes
here, and one can average over fields $B_{\mu}$ and $a_{\mu}$ independently
using the 'tHooft's identity\footnote{private communication to  the
author, December 1993.}
\be
 Z=\int
DA_{\mu} exp (-S(A)) = \frac{\int DB_{\mu}\eta(B)\int Da_{\mu} exp
(-S(B+a))}{\int DB_{\mu}\eta(B)}
  \ee
  $$ \equiv<<exp(-S(B+a)>_a>_B $$
   with
arbitrary weight $\eta(B)$. In our case we choose $\eta(B)$ to fix field
correlators and string tension at their observed values.

The partition function can be written as $$ Z(V,T,\mu=0) =<Z(B)>_B\;,$$
\begin{equation}
Z(B)=N\int D\phi exp (-\int^{\beta}_0 d\tau \int d^3x L(x,\tau))
\end{equation}
 where $\phi$ denotes all set of fields $a_{\mu}, \Psi, \Psi^+,N$ is a
normalization  constant, and the sign $<>_B$  means some averaging over
(nonperturbative) background fields $B_{\mu}$.

The inverse gluon propagator in the
background gauge is
\begin{equation}
W^{ab}_{\mu\nu} =- D^2(B)_{ab} \cdot \delta_{\mu\nu} - 2 g F^c_{\mu\nu}(B) f^{acb}
\end{equation}
where
\begin{equation}
(D_{\lambda})_{ca} = \partial_{\lambda} \delta_{ca} - ig T^b_{ca} B^b_{\lambda} \equiv
\partial_{\lambda} \delta_{ca} - g f_{bca} B^b_{\lambda}
\end{equation}

Integration over ghost and gluon degrees of freedom in (32) yields
\begin{eqnarray}
\nonumber
Z(B) =N'(det W(B))^{-1/2}_{reg} [det (-D_{\mu}(B) D_{\mu}(B+a))]_{a=\frac{\delta}{\delta J}} \times
\\
\times \{ 1+ \sum^{\infty}_{l=1} \frac{S_{int}^l}{l!} (a= \frac{\delta}{\delta J}) \}
exp (-\frac{1}{2} J W^{-1}J)_{J_{\mu}= \;\;\;\;\;D_{\mu}(B)F_{\mu\nu}(B)}
\end{eqnarray}

One can consider  strong background fields, so that $gB_{\mu}$ is large (as
compared to $\Lambda^2_{QCD}$), while $\alpha_s=\frac{g^2}{4\pi}$
in that strong background is small at all distances.

In this case Eq. (35) is a perturbative sum in powers of $g^n$,
arising from expansion in $(ga_{\mu})^n$.

In what follows we shall discuss the Feynman graphs for the free energy $F(T),$
connected to $Z(B)$ via
\begin{eqnarray}
F(T) = -T ln <Z(B)>_B
\end{eqnarray}

We are now in position to make expansion of $Z$ and $F$ in powers of
$ga_{\mu}$ (i.e. perturbative expansion in $\alpha_s$), and the
leading--nonperturbative term $Z_0, F_0$ -- can be represented as a sum of
contributions with different $N_c$ behaviour of which we systematically will
keep the leading terms $0(N_c^2),0(N_c)$ and $0(N_c^0)$.

To describe the temperature phase transition one should specify
phases and compute free energy. For the confining phase to lowest
order in $\alpha_s$ free energy is given by Eq.(36) plus contribution
of energy density $\varepsilon $ at zero temperature
\be
F(1)=\varepsilon V_3-\frac{\pi^2}{30}V_3T^4-T\sum_s\frac{V_3(2m_s
T)^{3/2}}{8\pi^{3/2}}e^{-m_{s/T}}+0(1/N_c)
\ee
where $\varepsilon$ is defined by scale anomaly [16]
\be
\varepsilon \simeq
-\frac{11}{3}N_c\frac{\alpha_s}{32\pi}<(F^a_{\mu\nu}(B))^2>
\ee
and the next terms in (37) correspond to the contribution of mesons (we
keep only pion gas) and glueballs. Note that $\varepsilon=0(N^2_c)$
while two other terms in (37) are $0(N^0_c)$.

For the second phase (to be the high temperature phase) we make an
assumption that there all color magnetic field correlators are the same as
in the first phase, while all color electric fields vanish. Since at
$T=0$ color--magnetic correlators (CMC) and color--electric
correlators (CEC) are equal due to the Euclidean $0(4)$ invariance,
one has
\be
<(F^a_{\mu\nu}(B))^2>=<(F^a_{\mu\nu})^2>_{el}+<(F^a_{\mu\nu})^2>_{magn};
<F^2>_{magn}=<F^2>_{el}
\ee

The string tension $\sigma$ which characterizes confinement is due to the
electric fields [5], e.g. in the plane (i4)
\be
\sigma=\sigma_E=\frac{g^2}{2}\int\int
d^2x<trE_i(x)\Phi(x,0)E_i(0)\phi(0,x)>+...
\ee
where dots imply higher order terms in $E_i$.

Vanishing of $\sigma_E$ liberates gluons and quarks, which will contribute
to the free energy in the deconfined phase their closed loop terms
 with all possible windings.
 As a result one
has for the high-temperature phase (phase 2).
 \be
F(2)=\frac{1}{2}\varepsilon
V_3-(N^2_c-1)V_3\frac{T^4\pi^2}{45}-\frac{7\pi^2}{180}N_cV_3T^4
n_f+0(N_c^0)
\ee

Comparing (37) and (41), $F(1)=F(2)$ at $T=T_c$, one finds in the
order $0(N_c)$, disregarding all meson and glueball contributions
\be
T_c=\left(\frac{\frac{11}{3}N_c\frac{\alpha_s<F^2>}{32\pi}}{\frac{2\pi^2}{45}
(N^2_c-1)+\frac{7\pi^2}{90}N_cn_f}\right)^{1/4}
\ee
For standard value of $G_2\equiv \frac{\alpha_s}{\pi}<F^2>=0.012
GeV^4$  (note that for $n_f=0$ one should use approximately 3
times larger value of $G_2$ ) one has for $SU(3)$ and different
values of $n_f=0,2,4$ respectively $T_c=~240,150,134$ MeV. This
should be compared with lattice data [16] $T_c(lattice)=240,146,131$
MeV.  Agreement is quite good.  Note that at large $N_c$ one has
$T_c=0(N_c^{0})$ i.e. the resulting value of $T_c$ doesn't depend on
$N_c$ in this limit. Hadron contributions to $T_c$ are $0(N_c^{-2})$
and therefore suppressed if $T_c $ is below the Hagedorn  temperature
as it typically happens in string theory estimates.

\section{Conclusion}

In all examples considered above gluon correlators and condensates entered as
a given input. It is important to find equations for FC which define them  up
to an overall scale to $\Lambda_{QCD}$. These equations  have been suggested
in [6] and are derived on the same physical basis as for the quark Green's
function in section 3. Namely one assumes that a gluon is propagating in the
nonperturbative background, described by FC, and one obtains  equations for
the gluon Green's function (in the field of the static charge, so that the
Green's function is gauge--invariant, as in the case of the quark in section
3).
\be
 (-\partial^2_\lambda \delta_{\mu\rho}
+\partial_{\mu}\partial_{\rho}) G_{\rho\nu}(x,y) +\int  M_{\mu\rho}^{(g)}
(x,z) G_{\rho\nu} (z,y) d^4z= \delta^{(4)}(x-y),
 \ee
where the mass operator $M^{(g)}$ is approximately equal at large distances
to $M^{(2,2)}$, where
\be
M^{(2,2)}_{\mu\nu}(x,y)=\frac{N_c}{C_2^f}\delta^{(4)}(x-y)
[J_{\lambda\lambda}
(x,y)\delta_{\mu\nu}-J_{\mu\nu}(x,y)],
\ee
For more extended treatment and derivation of equations (47),(48) the reader
is referred to [6].

The kernel $J_{\lambda\mu}$ in (44) is expressed through the field
correlator\\ $<FF>$. To make equations selfconsistent one should express the
latter through the gluon  Green's function $G$. This is possible  since
one can always refer the color indices in each term of $\sum^{N_c^2-1}_{a=1}
<F^aF^a>$ to the group of fields $b_\mu$. Then one can write symbolically
\be
tr<F(x)F(y)>\sim \partial_{\mu}\partial_{\nu} G(x,y) +(G(x,y))^2+perm+...
\ee
where ellipsis stands for higher cumulants.

The system of equations (43-44) allows for a nonperturbative solution, which
violates the scale invariance present in the equations. This solution is
defined by fixing  one nonperturbative scale, e.g. the string tension
$\sigma$. Then equations (43-44) predict that i) both field correlators
$D(x),D_1(x)$ [3] exponentially decay at large $x$; $D(x),D_1(x)\sim exp
(-x/T_g)$ in agreement with lattice data [4], and ii) the gluon correlation
length $T_g$ is connected to $\sigma$ as [6]
\be
1/T_g=(2.33)^{3/4}\sqrt{\frac{9\sigma}{2\pi}}
\ee
Insertion of the standard value $\sigma\approx 0.2 GeV^2$ yields $T_g\approx
0.2 fm$, which is in good agreement with lattice data [4].

We have derived equations for the gluon  and quark propagators in the field
of a static source. These equations possess symmetry (chiral for the quark and
scale  invariance for the gluon) which is  violated by the nonperturbative
solutions. One obtains in this way the CSB due to the confining kernel, and
the confining kernel itself satisfies nonlinear equations. Properties of this
kernel are in agreement with lattice data.

Thus the Gaussian model using the simplest FC can describe both qualitatively
and quantitatively the basic QCD phenomena at large $N_c$.

This work was supported by the RFFI grants 96-02-19184a, 96-1596740
and by RFFI-DFG grant 96-02-00088G.

\end{document}